\documentstyle[aps,epsf]{revtex}

\newcommand \beq {\begin{equation}}
\newcommand \eeq {\end{equation}}
\newcommand \boldsigma {\mbox{\boldmath $\sigma$}}
\newcommand \boldxi {\mbox{\boldmath $\xi$}}

\newcommand \rra {\rangle \rangle}
\newcommand \lla {\langle \langle}
\newcommand \bay {\begin{array}}
\newcommand \eay {\end{array}}

\begin{document}

\title{A multifractal phase-space analysis of perceptrons with biased patterns}
\author{ J. Berg \thanks{johannes.berg@physik.uni-magdeburg.de} and 
         A. Engel \thanks{andreas.engel@physik.uni-magdeburg.de}  \\
      {\small{\it Institut f\"ur Theoretische Physik,
      Otto-von-Guericke-Universit\"at Magdeburg}}\\
      {\small{\it PSF 4120,
      39016 Magdeburg, Germany}}}
 
\date{\small 13.1.98}
\maketitle

\begin{abstract}
We calculate the multifractal spectrum of the partition of the coupling
space of a perceptron induced by random input-output pairs with non-zero
mean. From 
the results we infer the influence of the input and output bias
respectively on both the storage and generalization properties of the
network. It turns out that the value of the input bias is irrelevant as long
as it is different from zero. The generalization problem with output bias
is new and shows an interesting two-level scenario. To compare our
analytical results with simulations we introduce a simple and efficient
algorithm to implement Gibbs learning. 

\end{abstract}


\section{Introduction}
The properties of simple networks of formal neurons 
can be quantitatively described by characterizing the partition of
the coupling space induced by the required input-output mappings
\cite{cover,Ga88,DGP,OpKi}. In some cases the geometrical properties of this
partition can be concisely specified by the multifractal spectrum of the
distribution of cells in coupling space corresponding to the different
output sequences that can be generated by the system for given inputs
\cite{MoKa,EnWe}. This approach has been used for a number of investigations
of both single-layer as well as multi-layer feed-forward neural nets
\cite{MoZe,BiOp,Cocco,Bex} and revealed a number of interesting new
aspects. The simplest case of the perceptron allows a 
rather detailed analysis also highlighting the problems and subtleties of
the method \cite{WeEn}. All investigations done so far have, however, 
considered symmetric statistics of both inputs and outputs. 

In the present paper we analyze the multifractal properties of the phase
space of a single-layer perceptron induced by input patterns with biased
statistics. A possible bias of the outputs is taken care of by considering
a special subset of cells only.
The investigation is motivated by the fact the storage properties of a
perceptron are known to depend markedly on the statistics of the patterns. A
similar influence can be expected for the generalization behaviour which has
to our knowledge not been studied for biased outputs before.

The analysis is performed using the methods that have been employed for the case
of unbiased patterns already. With the help of the replica trick the
multifractal spectrum $f(\alpha)$ averaged over the distribution of inputs
is calculated analytically for different pattern set sizes $\gamma$ . The
storage and generalization properties are determined by the points with
slope 0 and 1 respectively of these curves \cite{DGP,EnWe}. The results for
the storage capacity are
compared with previous findings \cite{Ga88}, whereas those for the
generalization behaviour are checked against numerical simulations. 
In order to efficiently explore the version space in these simulations, we 
introduce a randomized variant of the perceptron learning algorithm. 

\section{Calculation of the cell spectrum}
We consider a spherical perceptron specified by a real coupling vector
{\boldmath $J$} obeying $\sum_i J_i^2 =N$ and a set of $p=\gamma N$\ input
patterns $\boldxi^\mu$ with components $\xi_i^{\mu}$ drawn independently of
each other from the distribution
\beq\label{pdist}
P(\xi)=\frac{1-m}{2}\delta(\xi+1)+\frac{1+m}{2}\delta(\xi-1).
\eeq
To every input $\boldxi^\mu$ the perceptron determines an output
$\sigma^{\mu}$ according to 
\beq\label{perc}
\sigma^{\mu}=\mbox{sgn}(\frac{1}{\sqrt{N}}{\bf J}\cdot\boldxi^\mu)
            =\mbox{sgn}(\frac{1}{\sqrt{N}} \sum_{i=1}^{N} J_i \xi_i^{\mu})
\eeq
Every set of input patterns therefore divides the coupling space into $2^p$ 
cells 
\beq
\label{celldef}
C( \boldsigma ) = 
\{ {\bf J}|\;  \sigma^\mu = \mbox{sgn} ( {\bf J}\cdot\boldxi^\mu ) 
\;\forall \mu \}
\eeq
labelled by the sequence of outputs $\sigma^{\mu}$. Note that some
of the cells may be empty. The size $P(\boldsigma)=V(\boldsigma ) /$
$\sum_{{\mbox{\boldmath $\tau$}}}V({\mbox{\boldmath $\tau$} })$ of the cell gives the 
fraction of the coupling space that will produce the outputs $\sigma^{\mu}$\  
given the inputs $\xi^{\mu}_i$.

It is convenient to characterise the cell sizes by a \emph{crowding index} 
$\alpha (\boldsigma)$\ defined by     
\beq
P( \boldsigma ) = 2^{-N\alpha( \boldsigma)} \;.
\eeq
The entropy of the distribution of cell sizes in the thermodynamic limit 
averaged over the input patterns is then given by  
\beq
f(\alpha)=\lim_{N\to\infty} \frac{1}{N} \lla\log_2 
    {\sum_{\boldsigma}} \delta(\alpha-\alpha(\boldsigma))\rra
\eeq
where $\lla\ldots\rra$ denotes the average over the input pattern
distribution (\ref{pdist}).
Involving a trace over \emph{all} output sequences $\boldsigma$ this quantity
cannot be used to elucidate the dependencies on the output bias. In fact an
explicit calculation shows that it is also independent of the input pattern
distribution giving results for $f(\alpha)$ identical to
those for $m=0$. The intuitive reason for this fact is that a cell chosen at random 
from the above ensemble will with probability $1$ lie in the $N-1$\ dimensional 
subspace of the coupling space which is orthogonal to the direction of the bias. 
However the projection of the cell structure onto this subspace 
-- whose properties dominate the cell spectrum in the thermodynamic limit --  
carries no bias. 

Hence in order to study the influence of the input and output statistics on
the geometry of the phase space we have therefore restricted the
$\boldsigma$-trace to outputs with magnetization $m'$ according to
\beq
\label{distribution}
f(\alpha)=\lim_{N\to\infty} \frac{1}{N} \lla\log_2 
    {\sum_{\boldsigma}} ^{\prime} \delta(\alpha-\alpha(\boldsigma))\rra
         =\lim_{N\to\infty} \frac{1}{N} \lla\log_2 
    {\sum_{\boldsigma}} \delta(\frac{1}{\gamma N} \sum_{\mu=1}^{\gamma N}\sigma^{\mu} -m')
                          \delta(\alpha-\alpha(\boldsigma))\rra     
\;.     
\eeq
In the literature of multifractals $f(\alpha)$\ is called the 
\emph{multifractal spectrum}. It can be calculated by using its analogy 
with the microcanonical entropy of a spin system 
\boldsigma\ with Hamiltonian $\alpha(\boldsigma)$ and the free 
energy
\beq\label{freeenergy}
\tau ({q}) = - \lim_{N\to\infty} \frac{1}{N} \lla \log_2 Z \rra 
= - \lim_{N\to \infty} \frac{1}{N} \lla \log_2 {\sum_{\boldsigma}}^{\prime}
       P^{q} (\boldsigma )\rra
= - \lim_{N\to\infty} \frac{1}{N} \lla \log_2 {\sum_{\boldsigma}}^{\prime}
            2^{-{q}N\alpha(\boldsigma)} \rra
 \;.  
\eeq
The quenched average over input patterns with magnetisation $m$\ is 
performed using the pattern statistics (\ref{pdist}).
The entropy $f(\alpha)$\ can now be obtained by a Legendre-transformation with 
respect to the inverse temperature q
\beq\label{legendre}
f(\alpha) = \min_{{q}} [\alpha {q} - \tau({q})]\;.
\eeq

For the perceptron we have

\beq \label{size}
P(\boldsigma) = \int d\mu({\bf J}) \prod_{\mu =1}^p 
\theta( \frac{1}{\sqrt{N}}
\sigma^\mu {\bf J}\cdot \boldxi^\mu)\;.
\eeq
with the integral measure 
\beq
d\mu({\bf J}) = \prod_i \frac{dJ_i}{\sqrt{2\pi e}} \; \delta(N-{\bf J}^2) 
\eeq
ensuring the spherical constraint for the coupling vectors and the total 
normalisation over all cells $\sum_{\boldsigma} P(\boldsigma)=1$. 
$\theta(x)$\ is the Heaviside step function. In order to average $\log \sum P^{q}$ 
over the input patterns we introduce
two sets of replicas; one set labelled $a=1,\ldots ,n$\ for the standard 
replica-trick to replace the $\log$ and one set labelled $\alpha=1,\ldots ,q$ 
representing the $q$-th power of P in (\ref{freeenergy}). 
Thus we arrive at the replicated partition function

\begin{eqnarray}\label{Z^n}
Z_n & = &  \lla Z^n \rra \nonumber\\
  & = & \lla  \sum_{\{ \sigma_{\mu}^{a} \} } 
\prod_{a} \delta (\sum_{\mu}\sigma_{\mu}^{a}-N \gamma m')
\int\prod_{a,\alpha} d\mu ( {\bf J}_a^\alpha) \; 
\prod_{\mu,a,\alpha}
\theta(\frac{\sigma_\mu^a}{\sqrt{N}}{\bf J}_a^\alpha \cdot\boldxi^{\mu}) 
  \rra \;.
\end{eqnarray}

Averaging over the quenched disorder results in the order parameters 
\begin{eqnarray}\label{ops}
Q_{a,b}^{\alpha,\beta} &=& \frac{1}{N} {\bf J}_a^\alpha\cdot{\bf J}_b^\beta \nonumber \\
M_{a}^{\alpha} &=& \frac{1}{\sqrt{N}}\sum_{i}J_{i}^{a,\alpha}  \nonumber
\end{eqnarray}
as well as their conjugates $\hat{Q}_{a,b}^{\alpha,\beta}$\ and
$\hat{M}_{a}^{\alpha}$. In the
following $M_{a}^{\alpha}$\ will be referred to as the \emph{weak} magnetisation since
the mean value of the couplings -- the \emph{strong} magnetisation
$1/N \sum_{i}J_{i}^{a,\alpha}$ vanishes in the thermodynamic limit. The
appearance of an additional order parameter describing a ferromagnetic bias
of order $\sqrt{N}$ was to be expected. To produce biased outputs the local
fields ${\bf J}\cdot\boldxi^\mu/\sqrt{N}$ must have an average of order $1$.
Given $\lla \xi_i^{\mu}\rra = m$ this requires an average of the $J_i$ of
order $\sqrt{N}$. The integral representation of the delta-function
restricting the set of outputs introduces a further set of order parameters 
$R_{a}$. 

In the present paper we will only discuss the results obtained within the 
replica symmetric (RS) Ansatz \cite{MoKa,EnWe} 
\begin{eqnarray}\label{rs}
Q_{a,b}^{\alpha,\beta} & = &\left\{ \bay{ll} 1 & \;\; (a,\alpha)=(b,\beta)\\
                                             Q_1 & \;\; a=b, \alpha\neq\beta\\
                                             Q_0 & \;\; a\neq b \eay \right.  \nonumber\\
M_{a}^{\alpha} &=& M  \\
R_{a} &=& R \;. \nonumber 
\end{eqnarray}
This Ansatz represents the two-replica structure of the problem: $Q_1$\ is assigned 
to the overlap between coupling vectors belonging to \emph{identical cells} labelled by 
$\sigma^a_\mu$, $Q_0$\ is assigned to the overlap between \emph{different}
cells. For the cell spectrum without bias, it gives the correct result for $0\leq q \leq 1$, 
which is the interval of interest here. Nevertheless it remains plagued by divergences for 
$q<0$\ and continuous replica symmetry breaking for $q>1$, for details see \cite{WeEn}. 

Eliminating the conjugate order parameters one finds $Q_0=0$\ is always an extremum 
of $\tau(q)$ since for $Q_0=0$\ the saddle point equation in $Q_0$\ coincides with 
that in $M$. The interpretation of this result is that two randomly chosen 
coupling vectors with the same weak magnetisation do not have an overlap of order 1. 

We thus arrive at the free energy
\begin{eqnarray} \label{freeenergyres} 
\tau(q)& = & -\frac{1}{\ln2}{\mbox{extr}}_{Q_{1},M,R} \left[
                       \frac{q-1}{2}\ln(1-Q_1) + 
\frac{1}{2} \ln(1-(1-q)Q_1) \right. \nonumber\\
 &     &\left. - \gamma m' R 
                       +\gamma \ln (e^R\int Ds H^q_+ + e^{-R} \int Ds H^q_-) 
\right] \nonumber \\
H_+ &=& H(\frac{\sqrt{Q_1}s-mM}{\sqrt{1-Q_1}}) \\
H_- &=& H(\frac{-\sqrt{Q_1}s+mM}{\sqrt{1-Q_1}}) \nonumber 
\end{eqnarray}
where $Ds = \frac{ds}{\sqrt{2\pi}} \exp (-s^2/2)$\ is the Gaussian integral 
measure and $H(x)=\int_{x}^{\infty} Ds$. 

Extremising $\tau(q)$\ with respect to $Q_1$,$M$, and $R$ yields 
the saddle point equations
\begin{eqnarray} \label{spe}
\gamma \frac {\int Ds (e^R H_+^{q-2} +  e^{-R} H_-^{q-2}) G^2}
     {\int Ds (e^R H_+^{q} +  e^{-R} H_-^{q})} &=&  \frac{Q_1}{1-(1-q)Q_1} \nonumber \\
\frac {\int Ds (e^R H_+^{q-1} -  e^{-R} H_-^{q-1}) G}
     {\int Ds (e^R H_+^{q} +  e^{-R} H_-^{q})} &=& 0 \\
\frac {\int Ds (e^R H_+^{q} -  e^{-R} H_-^{q})} 
   {\int Ds (e^R H_+^{q} +  e^{-R} H_-^{q})} &=& m'\nonumber
\end{eqnarray}
where $G=1/\sqrt{2 \pi} \exp (-\frac{(\sqrt{Q_1}s-mM)^2}{2(1-Q_1)})$. 

The cell spectrum $f(\alpha)$\ can now be evaluated using (\ref{legendre}). 

\section{Discussion} 
For $m'=0$ the cell spectrum of unbiased patterns is recovered for any $m$. 
For $m=0$ no coupling vector with a weak magnetisation which produces a sequence of 
outputs with $m'\neq 0$ exists. We thus turn to the case  $m'\neq 0, m\neq 0$. 
A transformation $mM \rightarrow M$ in the free energy (\ref{freeenergyres}) 
would remove the magnetisation of the inputs from the saddle point equations 
(\ref{spe}). Thus the properties of the
cell spectrum for non-zero $m$\ and $m'$\ do not depend on $m$\ but only on 
$\gamma$\ and $m'$\ whereas the weak magnetisation of the couplings $M$\ scales 
with $m^{-1}$ for fixed $m'$\ and $\gamma$. Hence we may restrict the discussion 
to the case $m=m' \neq 0$\ without loss of generality. 

\begin{figure}[htb]
 \label{fig_spect} 
 \epsfysize=10cm
      \epsffile{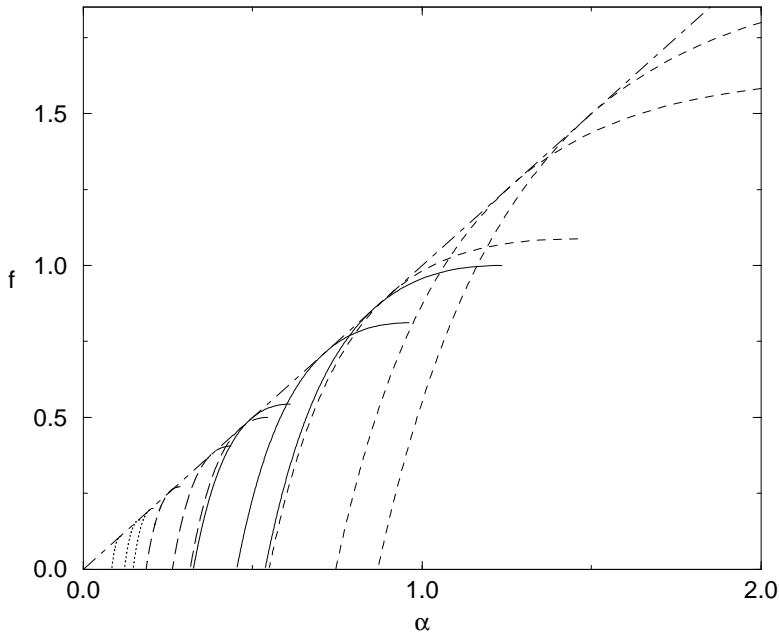}
\caption
{The multifractal spectrum $f(\alpha)$\ for various values of the loading parameter 
$\gamma=$0.2 (dotted), 0.5 (long dashed), 1 (full), and 2 (dashed) and values of the 
magnetisation $m'=0,0.5,0.75$\ from top to bottom respectively. The parts
with negative slope have been omitted since their interpretation is presently 
not clear (cf. [11]).}
\end{figure}  

Figure 1 shows the cell spectrum at several loading capacities $\gamma$\ and
magnetisations $m'$. For any given $q\equiv df / d \alpha$\ 
the number of cells decreases exponentially with increasing $m'$.
This is in accordance with the fact that the maximal possible number of
cells with output bias $m'$ scales as $2^{Nf_{tot}}$\ with
$f_{tot}=\gamma ((1-m')/2 \log_2 (1-m')/2 + (1+m')/2 \log_2 (1+m')/2)$.
The maximum $f_{max}$ of $f(\alpha)$\ exponentially dominates the \emph{number of cells}
${\mathcal{N}}=\int d \alpha 2^{N f(\alpha)}$,
hence a randomly chosen output sequence will result in a cell of size $\alpha(q=0)$, which
is termed a \emph{storage cell}.
For values of the loading parameter below the critical storage capacity
$\gamma_c$ typically all possible cells may be realised, so
$f_{max}=f_{tot}$. However above the critical storage capacity
only an exponentially small fraction of all possible cells may be realised
hence $f_{max}<f_{tot}$. Note that although $f_{max}$ decreases with increasing $m'$
as shown in figure 1 it does so slower than $f_{tot}$ so that the storage
capacity \emph{increases}. This can also be seen 
by comparing the curves of $f(\alpha)$\ at $\gamma=2$ for
$m'=0$ and $m'=.75$. The maximum of $f(\alpha)$\ 
is attained for $m'=0$ as $\alpha \rightarrow \infty$  indicating the cell volume 
shrinks to zero which signals the critical storage capacity. 
However for $m'=.75$\ the maximum of $f(\alpha)$\ is reached at 
finite values of $\alpha$\ indicating a finite size of a storage cell at $\gamma=2$. 
In fact the limit $q \rightarrow 0$\ of (\ref{freeenergyres}) can be taken explicitly 
yielding the saddle point equations for the storage problem for magnetised 
patterns \cite{Ga88}. 

On the other hand the cells dominating the \emph{volume}\ 
${\mathcal{V}}=\int d \alpha 2^{N (f(\alpha)-\alpha)}$\ 
of the coupling space are characterised by $q\equiv \frac{df}{d\alpha}=1$. In general 
such a cell is taken to describe the generalisation behaviour of the perceptron, since 
in the thermodynamic limit a randomly chosen teacher perceptron 
will lie within a cell of this size with probability one. 
The saddle point equations (\ref{spe}) at this point are 
\begin{eqnarray} 
Q_1 &=& \gamma \int Ds (H^{-1}_+ + H^{-1}_-)G^2  \label{spegen1}\\
R &=& 0 \label{spegen3}\\
m' &=& 1-2H(mM) \label{spegen2}\;.
\end{eqnarray}
The interpretation of this saddle point may not be immediately obvious,
since we have specified the magnetisation of the outputs, but no
properties of a teacher that will produce such a set of outputs
on a given input pattern. However equation (\ref{spegen2})
provides an explicit relation between $m$, $m'$, and $M$.
It gives the weak magnetisation $M$\ of the couplings of a teacher or
student required to produce a magnetisation $m'$\ of the outputs 
given a set of inputs with magnetisation $m$. In the context of 
a teacher with magnetisation $M$\ acting on a set of inputs with 
magnetisation $m$\ (\ref{spegen2}) simply follows from the central 
limit theorem and $Q_1$\ describes the overlap between teacher 
and student after the student has learned to classify $\gamma N$ 
examples correctly. The subsequent generalisation error $\epsilon_g$\ 
is given by
\beq \label{generr}
\epsilon_g=1 - \frac{2}{\sqrt{1-Q_1^2}}\int^{\infty}_0 \int^{\infty}_0 
               \frac{d\chi d\sigma}{2\pi}
               \exp(-\frac{\sigma^2-2\sigma \chi Q_1 + \chi^2 + 2 m^2 M^2 (1-Q_1)}{2(1-Q_1^2)})
               \cosh(m M \frac{\chi + \sigma}{1+Q_1}) 
\eeq

\begin{figure}[htb]
 \epsfysize=10cm
      \epsffile{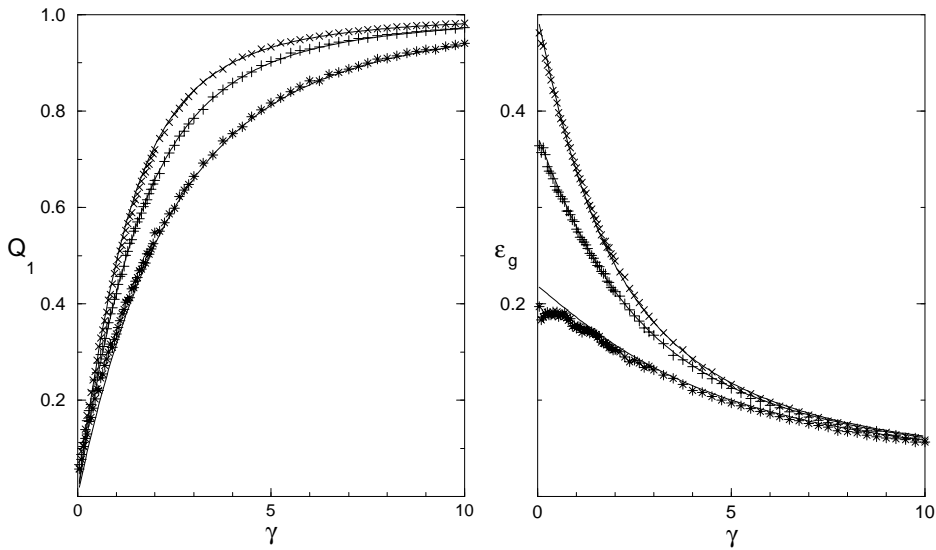}
\caption{The teacher-student overlap $Q_1$\ and the resulting 
generalisation error $\epsilon_g$\ as a function of $\gamma$\ for $m'=0,0.5,0.75$\ 
(from top to bottom). The uppermost curve corresponds to the results of 
Gy\"orgyi and Tishby [12]. The full lines are analytical curves 
whereas the symbols are the results of numerical simulations with $N=200$\ 
averaged over 200 patterns. The symbol size corresponds to 5 times 
the statistical error.}
\end{figure}  

The full lines in fig.2 show the overlap $Q_1$\
between student and teacher and the corresponding generalisation
error $\epsilon_g$\ as a function of $\gamma$\ after the student has learned
$\gamma N$\ examples for different magnetisations $m'=m$.
The generalisation error is found to decrease  for increasing $m'$\
at fixed $\gamma$.
The effect is most pronounced at low values of $\gamma$; in particular we
find $\epsilon_g(\gamma=0,m')=\frac{1-m'^2}{2}$. In fact the
weak magnetisation of the couplings $M$\ is independent of the loading parameter and
already takes on its non-zero value (for $m\neq 0$) at $\gamma=0$. However
two independent output strings $\sigma_T^{\mu}$ and $\sigma_J^{\mu}$ with
the same average $m'$ differ on average in $(1-m'^2)/2$ bits only. Hence
$\epsilon_g < .5$ even for zero teacher-student overlap $Q_1$. Qualitatively
this means that the student learns the correct \emph{bias} of the outputs
after a {\em non-extensive number} of examples already. 
A  plausibility  argument underlining this effect is a follows: 
By the central limit theorem the sum $1/\sqrt{N} \sum_{i} J_i \xi_i$\ is a 
Gaussian variable with mean $mM$\ and variance $1$. Hence equation 
(\ref{spegen2}) needs to be fulfilled if an output of the same sign as $m'$ 
is to be produced with probability $(1+m')/2$. If the number of examples $p$ 
becomes infinite in the thermodynamic limit the number of outputs of the same sign 
as $m'$ tends to $p$ times this probability. 
This implies that it is sufficient for the number of examples to scale 
with $N^\delta$, $0<\delta<1$, so the number of examples is infinite in 
the thermodynamic limit, for $M$\ to take on its saddle point value.   

\section{Simulation algorithm and numerical results}
Gibbs learning at $T=0$\ is a convenient tool for the analytical study of 
generalisation problems, since it characterizes the typical performance of 
a compatible student. It is however not completely straightforward to 
implement in numerical simulations. The necessary average over the 
compatible students cannot be performed directly because the version space is only an 
exponential fraction of the high-dimensional coupling space of the perceptron. 
Several methods to circumvent this problem have been suggested, including 
a random walk in the version space of the student \cite{ranwalk}, a billiard in 
version space \cite{billiard}, or a variant of the Adatron algorithm, where in 
each realisation a few randomly chosen patterns are learnt in addition 
to the examples \cite{ranada}. 

Here we propose the {\it randomized perceptron algorithm}\ as a new method to 
effectively simulate Gibbs learning and apply it to the specific problem of 
biased patterns. Starting with all couplings equal to zero, the randomized 
perceptron algorithm runs over all examples, leaving the couplings unchanged if the 
pattern is classified correctly by their present values.    
If a pattern is not classified correctly, the 
algorithm adds the standard Hebbian term as well as a random vector 
with components chosen independently from a Gaussian distribution. 
Since for large $N$\ the random vectors are all orthogonal to each 
other, the coupling vector will end up in version space even though 
the amplitudes of the Hebbian and the random terms are of comparable magnitude. 
This procedure slows down the convergence of the perceptron 
algorithm but leads to more reliable results for the generalisation error. 
The standard deviation of the noise term was 
$4.8 \gamma$\ (compared to the magnitude of the Hebbian term of $2 \gamma$), 
but no strong dependence on the standard deviation was observed on the 
interval $2 \gamma$ to $10 \gamma$. 

The simulations whose results are shown in Fig. 2 were performed with a system size 
of $N=200$. Sets of Gaussian distributed 
inputs with magnetisation $m$\ were generated and the components of the couplings of 
potential teacher perceptrons were chosen from a Gaussian distribution with zero mean. 
In this way each teacher perceptron was given a weak magnetisation. Teachers were generated 
until one of them produced an output magnetisation $m'$\ on the given inputs. 
This teacher was used to generate the outputs used in the subsequent step:  
The resulting patterns were taught to the student using the randomized perceptron 
algorithm and its overlap with the teacher and the generalisation error were evaluated.      

Except at large values of the magnetisation $m'$, where finite size effects 
are more noticeable, the numerical results are in very good agreement with the 
analytical curves. 

\section{Summary}
In the present paper we have investigated the influence of a bias in the
distribution of inputs and outputs on the cell structure in the phase
space of a perceptron. To this end the multifractal spectrum $f(\alpha)$ was
calculated analytically for different values of the storage ratio $\gamma$
with the help of the methods used already for the case of unbiased patterns.
Both the storage and the generalization properties can be read off from
the behaviour of $f(\alpha)$. 

For the storage problem we showed that the input bias has no influence on 
the storage capacity provided the output bias $m'$ is equal to
zero. If both input and output bias are non-zero the storage capacity
$\gamma_c$ increases with increasing output bias $m'$ irrespectively of the
value of the input bias. Biased patterns have been considered before in the context 
of the phase space volume of attractor neural networks \cite{Ga88}. In this case it 
is natural to assume $m=m'$. Our results show that for the perceptron this case is 
in fact generic as for non-zero $m$ the properties of the entire cell spectrum 
only depend on $m'$. The case $m=0$ but $m' \neq 0$ cannot be realized by a perceptron 
with weak magnetisation of the couplings without thresholds \cite{BiehlOpp}\ 
and thus was not treated here. 
The behaviour of the maximum $f_{max}$
of $f(\alpha)$ as a function of $m'$ generalize the results about the
number of dichomoties \cite{cover} to $m'\neq 0$.

For the generalization problem we found an interesting two-level scenario of
learning. The student first determines the weak magnetization of the teacher couplings
necessary to produce outputs of the required bias. This is accomplished
after a non-extensive number of training examples already. The curves for
the generalization error therefore start off at $\gamma=0$ with values
smaller than $.5$. In the second step the student then reduces the
generalization error further in the usual way. The asymptotic behaviour is
not modified by the bias of the patterns. The analytical results are in
excellent agreement with numerical simulations. We have found a randomized
variant of the perceptron algorithm a simple device for reliable simulations of
Gibbs learning. 

Biased patterns can be viewed as the
simplest example of a hierarchy of inputs. It would be interesting to see
whether the generalization strategy observed  can also be found in the
general case of hierarchically correlated patterns in the sense that the student
first learns the classes of patterns and only then the individual
representatives.\\[.2cm]

{\bf Acknowledgments: }Many thanks to Martin Weigt for stimulating discussions 
and critical reading of the manuscript. JB gratefully acknowledges financial 
support by the \emph{Studienstiftung des Deutschen Volkes}.

\end{document}